\documentclass[aps,prl,twocolumn,superscriptaddress,longbibliography]{revtex4-1}
\usepackage{enumerate}
\usepackage{graphicx}
\usepackage{amsmath}
\usepackage{mathtools}
\usepackage{xcolor}
\usepackage{amsfonts}
\usepackage{amssymb}
\usepackage{units}
\usepackage[normalem]{ulem}
\newcommand{\ket}[1]{| #1 \rangle}

\begin{document}
\today
\title{Hierarchy of magnon entanglement in antiferromagnets}

\author{Vahid Azimi Mousolou\footnote{Electronic address: v.azimi@sci.ui.ac.ir}}
\affiliation{Department of Physics and Astronomy, Uppsala University, Box 516, 
SE-751 20 Uppsala, Sweden}
\affiliation{Department of Applied Mathematics and Computer Science, 
Faculty of Mathematics and Statistics, 
University of Isfahan, Isfahan 81746-73441, Iran}

\author{Andrey Bagrov}
\affiliation{Department of Physics and Astronomy, Uppsala University, Box 516, 
SE-751 20 Uppsala, Sweden}

\author{Anders Bergman}
\affiliation{Department of Physics and Astronomy, Uppsala University, Box 516, 
SE-751 20 Uppsala, Sweden}

\author{Anna Delin }
\affiliation{Department of Physics and Astronomy, Uppsala University, Box 516, 
SE-751 20 Uppsala, Sweden}
\affiliation{Department of Applied Physics, School of Engineering Sciences, 
KTH Royal Institute of Technology, AlbaNova University Center, SE-10691 Stockholm, 
Sweden}
\affiliation{Swedish e-Science Research Center (SeRC), KTH Royal Institute of Technology, 
SE-10044 Stockholm, Sweden}

\author{Olle Eriksson}
\affiliation{Department of Physics and Astronomy, Uppsala University, Box 516, 
SE-751 20 Uppsala, Sweden}
\affiliation{School of Science and Technology, \"Orebro University, SE-701 82, 
\"Orebro, Sweden}

\author{Yuefei Liu}
\affiliation{Department of Applied Physics, School of Engineering Sciences, 
KTH Royal Institute of Technology, AlbaNova University Center, SE-10691 Stockholm, 
Sweden}

\author{Manuel Pereiro}
\affiliation{Department of Physics and Astronomy, Uppsala University, Box 516, 
SE-751 20 Uppsala, Sweden}

\author{Danny Thonig}
\affiliation{School of Science and Technology, \"Orebro University, SE-701 82, 
\"Orebro, Sweden}

\author{Erik Sj\"oqvist\footnote{Electronic address: 
erik.sjoqvist@physics.uu.se}}
\affiliation{Department of Physics and Astronomy, Uppsala University, 
Box 516, SE-751 20 Uppsala, Sweden}

\begin{abstract}
Continuous variable entanglement between magnon modes in Heisenberg antiferromagnet 
with Dzyaloshinskii-Moryia (DM) interaction is examined. Different bosonic modes are 
identified, which allows to establish a hierarchy of magnon entanglement in the ground 
state. We argue that entanglement between magnon modes is determined by a simple 
lattice specific factor, together with the ratio of the strengths of the DM and Heisenberg 
exchange interactions, and that magnon entanglement can be detected 
by means of quantum homodyne techniques. As an illustration of the relevance of our 
findings for possible  entanglement experiments in the solid state, a typical antiferromagnet 
with the perovskite crystal structure is considered, and it is shown that long wave length 
magnon modes have the highest degree of entanglement. 
\end{abstract}
\maketitle
Quantum entanglement allows particles to act as a single non-separable entity, no matter 
how far apart they are. This is the feature that was initially used in the Einstein-€"Podolsky-€"Rosen 
(EPR) argument against completeness of quantum 
mechanics 
\cite{einstein35}. The original form of the EPR argument is closely related 
to continuous variable (CV) entanglement 
\cite{ou1992, giedke03, reid09}, which describes entanglement between bosonic modes. 
Such systems are characterized by an infinite number of allowed states, which makes 
them very different from the finite-dimensional Hilbert spaces associated with discrete 
variable (e.g., qubit) systems. Nevertheless, just as discrete variable entanglement, CV 
entanglement provides an essential resource for quantum technologies allowing for universal 
quantum information processing \cite{braunstein2005}, the realization of quantum 
teleportation \cite{braunstein2005, braunstein1998, Opatrny2001}, quantum memories 
\cite{braunstein2005, hammerer2010}, and quantum enhanced measurement resolution 
\cite{giovannetti2004}.

It is natural to expect that quantum systems, in which information is carried in a wave-like 
form, in general can show entanglement. However, the question is how clear the entanglement 
can be demonstrated and what quantum systems might be appropriate for potential 
applications. In the solid state, there are several collective modes that could be suitable 
hosts of entanglement. Here, we focus on magnons, collective wave-like excitations of 
a magnet with a well-established quantum nature \cite{mohn2006}. Typically magnons 
can be found in energy range of up to $\sim$ 500 meV, and with wave lengths spanning 
a range of hundreds of lattice constants to just a few. Low energy magnon excitations 
can be observed in different classes of magnetic materials, such as ferromagnets, 
antiferromagnets, and ferrimagnets, and each class has a vast space of materials to 
choose from. One of the broadest 
classes of antiferromagnets can be found in oxide compounds, in particular transition 
metal oxides \cite{googenough2020}, where for long wave lengths the dispersion relation 
is essentially linear. 

In this paper, we focus on magnon CV entanglement in antiferromagnets, in 
which both the Heisenberg exchange and the Dzyaloshinskii-Moriya (DM) interactions 
may be relevant \cite{mohn2006}. To begin with, we consider the quantum antiferromagnetic 
Heisenberg Hamiltonian on a bipartite lattice
\begin{eqnarray}
H_{0}=J\sum_{\langle ij\rangle}{\bf S}_{i}\cdot {\bf S}_{j}, \ \ \ J>0,
\label{MH}
\end{eqnarray}
where ${\bf S}_{j}$ is the spin-$S$ operator on site $j$ and $J$ is 
the strength of the exchange interaction. 
Using the Holstein-Primakoff (HP) transformation at low 
temperatures ($k_{B}T\ll J$) followed by the Fourier 
transformation, one can express the spin Hamiltonian (trivial terms and zero-point 
energies are neglected from now on) in terms of bosonic operators as 
\begin{eqnarray}
H_{0} = zJS \sum_{\mathbf{k}} 
\left[ a_{\mathbf{k}}^{\dagger}a_{\mathbf{k}} + 
b_{\mathbf{k}}^{\dagger} b_{\mathbf{k}} 
+ \gamma_{\mathbf{k}}a_{\mathbf{k}} 
b_{\mathbf{k}}+\gamma_{-\mathbf{k}}a_{\mathbf{k}}^{\dagger} 
b_{\mathbf{k}}^{\dagger}\right]  
\label{MHM}
\end{eqnarray}
with the lattice specific parameter $\gamma_{\mathbf{k}}=\frac{1}{z} 
\sum_{\mathbf{ \mathbf{\delta}}}e^{i\mathbf{k} \cdot 
\boldsymbol{\delta}}$, $z$ being the coordination number of the 
lattice and the sum over $\boldsymbol{\delta}$ is carried out over 
nearest neighbors. Here, $a_{\mathbf{k}}^{\dagger}$ ($a_{\mathbf{k}}$) 
and $b_{\mathbf{k}}^{\dagger}$ ($b_{\mathbf{k}}$) are bosonic creation 
(annihilation) operators representing two magnon modes with wave vector 
$\mathbf{k}$ that are associated with the two sublattices (see Supplemental Material).  

By employing the Bogoliubov transformation 
 \begin{eqnarray}
\left(
\begin{array}{cc}
  a_{\mathbf{k}}    \\
   b_{\mathbf{k}}^{\dagger}       
\end{array}
\right)=\left(
\begin{array}{cc}
  u_{\mathbf{k}}& v_{\mathbf{k}}    \\
v_{\mathbf{k}}^{*}& u^{*}_{\mathbf{k}}       
\end{array}
\right)\left(
\begin{array}{cc}
  \alpha_{\mathbf{k}}    \\
   \beta_{\mathbf{k}}^{\dagger}       
\end{array}
\right),
\label{eq:FBT}
\end{eqnarray}
with $u_{\mathbf{k}}$ and $v_{\mathbf{k}}$ given by 
\begin{eqnarray}
|u_{\mathbf{k}}|^{2} & = & 
\frac{1}{2\sqrt{1-|\gamma_{\mathbf{k}}|^{2}}} + \frac{1}{2} , 
 \ \ 
|v_{\mathbf{k}}|^{2} = 
\frac{1}{2\sqrt{1-|\gamma_{\mathbf{k}}|^{2}}} - \frac{1}{2},
\nonumber \\ 
 & & \frac{v_{\mathbf{k}}}{u^{*}_{\mathbf{k}}} = -  
\frac{1-\sqrt{1-|\gamma_{\mathbf{k}}|^{2}}}{\gamma_{\mathbf{k}}},
\label{eq:bogoheisenberg}
\end{eqnarray}
where $|\gamma_{\mathbf{k}}|<1$, we obtain the Hamiltonian in 
diagonal form 
 \begin{eqnarray}
H_{0} = \sum_{\mathbf{k}} \epsilon_{\mathbf{k}}(\alpha_{\mathbf{k}}^{\dagger}
\alpha_{\mathbf{k}}+\beta_{\mathbf{k}}^{\dagger} \beta_{\mathbf{k}}) 
\label{MHM2}
\end{eqnarray}
in terms of the new bosonic operators $(\alpha,\beta)$. 
For the 
antiferromagnetic magnon dispersion 
relation, we find $\epsilon_{\mathbf{k}} = 
zSJ\sqrt{(1-|\gamma_\mathbf{k}|^{2})}$ (see Supplemental Material for derivation). 
 
\begin{figure}[h]
\begin{center}
\includegraphics[width=70mm]{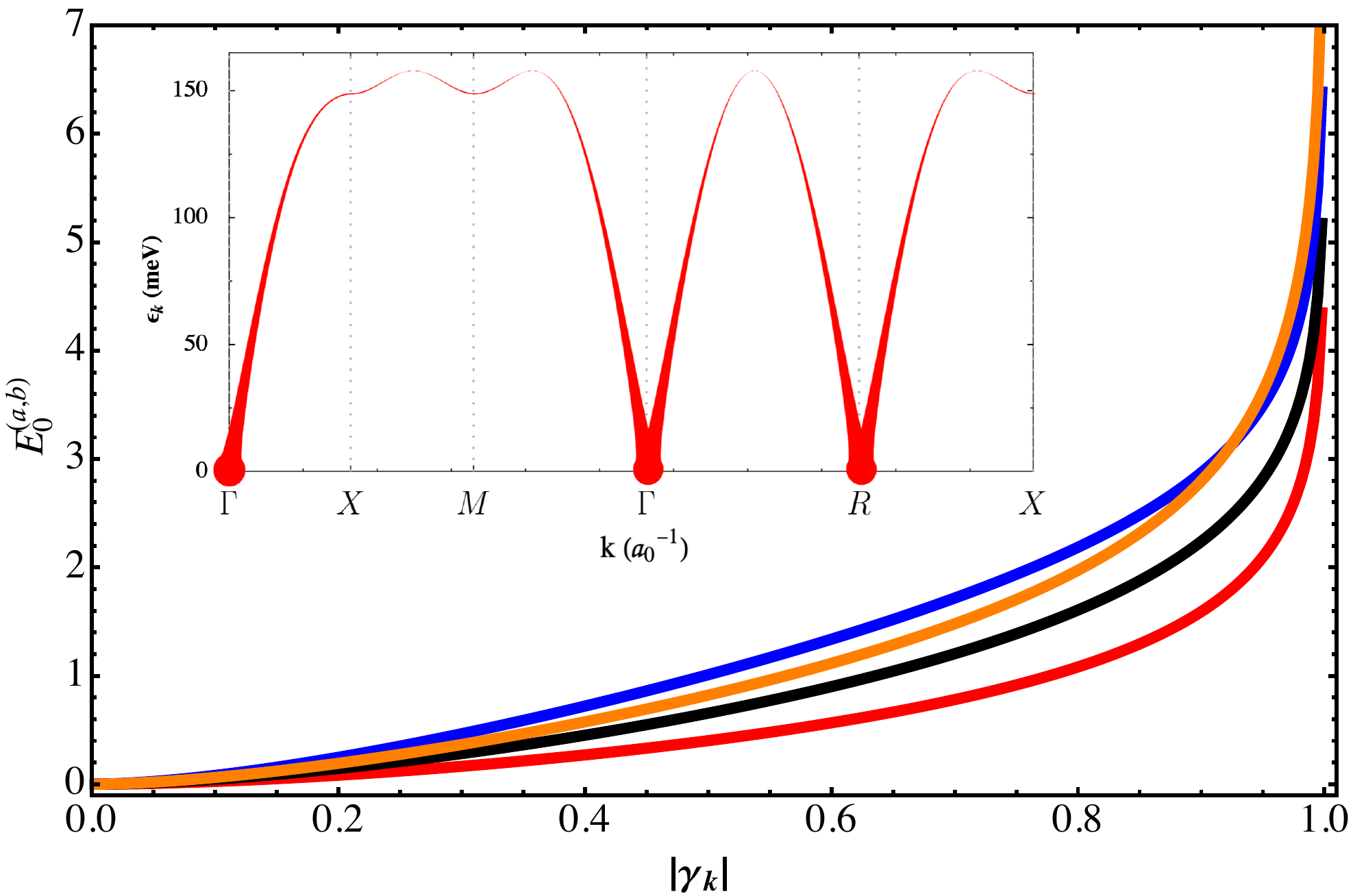}
\end{center}
\caption{(Color online). The entropy of entanglement 
$E_{0}^{(a, b)}$ for the two-€"mode generalized coherent state 
$\ket{r_{\mathbf{k}}, \phi_{\mathbf{k}}}$ (red curve) and for the 
corresponding first and second excited states as a function of $\left| \gamma_\mathbf{k} \right|$. 
The magnon CV entanglement in the 
first excited states ($\alpha_{\mathbf{k}}^{\dagger}\ket{r_{\mathbf{k}}, \phi_{\mathbf{k}}}$ 
and $\beta_{\mathbf{k}}^{\dagger}\ket{r_{\mathbf{k}}, 
\phi_{\mathbf{k}}}$) is shown in black. Blue and orange curves illustrate 
$E_{0}^{(a, b)}$ in the second exited states 
$(\alpha_{\mathbf{k}}^{\dagger})^{2}\ket{r_{\mathbf{k}}, \phi_{\mathbf{k}}}$ 
(or $(\beta_{\mathbf{k}}^{\dagger})^{2} \ket{r_{\mathbf{k}}, 
\phi_{\mathbf{k}}}$) and 
$\alpha_{\mathbf{k}}^{\dagger}\beta_{\mathbf{k}}^{\dagger}\ket{r_{\mathbf{k}}, 
\phi_{\mathbf{k}}}$, respectively. The inset depicts magnon dispersion of SrMnO$_3$ for 
a selected path of $\mathbf{k}$ along high-symmetry directions of the BZ. The width of 
the bands depicts the entropy of entanglement.}
\label{fig:HGSE}
\end{figure}

The ground state of $H_{0}$ in the $(\alpha,\beta)$ modes reads 
$\ket{\psi_{0}} = \prod_{\mathbf{k}} \ket{0; \alpha_{\mathbf{k}}} 
\ket{0;\beta_{\mathbf{k}}}$, which is a separable vacuum state with 
vanishing entropy of entanglement \cite{remark1}, i.e., 
$E_{0}^{(\alpha,\beta)}=0$.  
By making the inverse Bogoliubov transformation, 
we may express the ground state as  
\begin{eqnarray}
\ket{\psi_{0}}=\prod_{\mathbf{k}} \ket{r_{\mathbf{k}}, 
\phi_{\mathbf{k}}}   
\end{eqnarray}
with the two-mode 
generalized coherent state 
\begin{eqnarray}
\ket{r_{\mathbf{k}}, \phi_{\mathbf{k}}}=\frac{1}{\cosh 
r_{\mathbf{k}}}\sum_{n=0}^{\infty} 
e^{in\phi_{\mathbf{k}}}\tanh^{n}r_{\mathbf{k}} 
\ket{n; a_{\mathbf{k}}}\ket{n; b_{\mathbf{k}}} 
\end{eqnarray}
in the $(a,b)$ occupation number basis $\ket{n; a_{\mathbf{k}}}$ and 
$\ket{n;b_{\mathbf{k}}}$ (see Supplemental Material for derivation). 
Here, the parameter $r_{\mathbf{k}}$ and the phase $\phi_{\mathbf{k}}$ 
are given by $e^{i\phi_{\mathbf{k}}} \tanh 
r_{\mathbf{k}}=\frac{v_{\mathbf{k}}}{u^{*}_{\mathbf{k}}}$, 
$r_{\mathbf{k}} \equiv r_{\mathbf{k}} (|\gamma_{\bf k}|) 
\geqslant 0$ and $\phi_{\mathbf{k}} \equiv 
\phi_{\mathbf{k}} (\gamma_{\bf k}) = \pi - 
\arg[\gamma_\mathbf{k}]$. The entropy of entanglement 
for $\ket{r_{\mathbf{k}}, \phi_{\mathbf{k}}}$ 
\cite{giedke03, remark1} 
\begin{eqnarray}
E_{0}^{(a, b)}&=&\cosh^{2}r_{\mathbf{k}} \log_2 
\left[ \cosh^{2}r_{\mathbf{k}} \right]-\sinh^{2}r_{\mathbf{k}} \log_2 
\left[ \sinh^{2}r_{\mathbf{k}} \right] 
\nonumber\\
 & = & |u_\mathbf{k}|^{2} \log_2
|u_\mathbf{k}|^{2} - |v_\mathbf{k}|^{2} \log_2 |v_\mathbf{k}|^{2},
\label{EEH0}
\end{eqnarray}
evaluates CV entanglement between two magnon modes $a_{\mathbf{k}}$ and $b_{\mathbf{k}}$.
This expression indicates that the magnon CV entanglement in 
the $(a, b)$ modes is, in the low temperature regime, solely determined 
by the lattice geometry encoded in the $\gamma_\mathbf{k}$ parameter. 
Fig.~\ref{fig:HGSE} shows how the 
entropy of entanglement for the two-€"mode generalized coherent state
varies with $|\gamma_\mathbf{k}|$, and the inset shows its dependence 
on $\mathbf{k}$. The analysis presented here is appropriate for many classes of 
compounds. A concrete example that is known to exhibit only nearest neighbor Heisenberg 
exchange is SrMnO$_3$ with $J=\unit[17.1]{meV}$ \cite{zhu2020}. The magnon dispersion 
of this magnetic insulator is known, both from experiments and theory, and it is shown in 
Fig.~\ref{fig:HGSE} (inset), where the band width illustrates the entropy of entanglement as 
a function of ${\bf k}$. As is clear from the figure, when $|\gamma_\mathbf{k}|$ 
approaches 1, the two-mode magnon entanglement becomes stronger and the entropy 
of entanglement formally diverges. The fact that the entanglement is largest close to the 
zone center is important since magnons typically are more distinct and 
long-lived in this regime, in comparison to the more short wave length 
magnons that have higher damping \cite{mohn2006}.

Although the main focus of our study is on the ground state, we also consider 
magnon CV entanglement for the two-fold degenerate first excited states 
$\alpha_{\mathbf{k}}^{\dagger}\ket{r_{\mathbf{k}}, \phi_{\mathbf{k}}}$ and 
$\beta_{\mathbf{k}}^{\dagger} \ket{r_{\mathbf{k}}, \phi_{\mathbf{k}}}$, as well as 
for the three-fold degenerate second excited states  
$(\alpha_{\mathbf{k}}^{\dagger})^{2}\ket{r_{\mathbf{k}}, \phi_{\mathbf{k}}}$, 
$(\beta_{\mathbf{k}}^{\dagger})^{2} \ket{r_{\mathbf{k}},\phi_{\mathbf{k}}}$, and 
$\alpha_{\mathbf{k}}^{\dagger} \beta_{\mathbf{k}}^{\dagger} 
\ket{r_{\mathbf{k}}, \phi_{\mathbf{k}}}$. For these states, 
we find that the entropy of entanglement behaves in a similar way as 
for the ground state, see Fig.~\ref{fig:HGSE}, though being slightly larger.

The arithmetic mean of the squared quadrature variances is directly related to the 
geometry of the spin lattice
\begin{eqnarray}
\Delta(r_{\mathbf{k}},  \phi_{\mathbf{k}}) & = &  
\frac{1}{2}[\text{Var}^{2}_{r_{\mathbf{k}}, 
\phi_{\mathbf{k}}}(X_{\mathbf{k}}^{A}-X_{\mathbf{k}}^{B})
 + \text{Var}^{2}_{r_{\mathbf{k}},
\phi_{\mathbf{k}}}(P_{\mathbf{k}}^{A}+P_{\mathbf{k}}^{B})]
\nonumber \\ 
 & = & \cosh 2r_{\mathbf{k}}-\sinh
2r_{\mathbf{k}}\cos\phi_{\mathbf{k}} = 
\frac{ 1+\mathrm{Re}[\gamma_{\mathbf{k}}]}{\sqrt{1-|\gamma_{\mathbf{k}}|^{2}}} 
\label{AMV}
\end{eqnarray}
with $\mathrm{Re}[\gamma_{\mathbf{k}}]$ being the real part of $\gamma_{\mathbf{k}}$. 
Here, $X_{\mathbf{k}}^{A} (X_{\mathbf{k}}^{B})$ and $P_{\mathbf{k}}^{A} (P_{\mathbf{k}}^{B})$ 
are the dimensionless position and momentum quadratures of the 
$a_{\mathbf{k}} (b_{\mathbf{k}})$ mode \cite{note1}, and 
$\text{Var}_{r_{\mathbf{k}}, \phi_{\mathbf{k}}}(V)$ is the variance of a given 
Hermitian operator $V$ with respect to the state $\ket{r_{\mathbf{k}}, \phi_{\mathbf{k}}}$. 

For $\Delta(r_{\mathbf{k}} ,  \phi_{\mathbf{k}})<1$, which corresponds to 
$\mathrm{Re}[\gamma_{\mathbf{k}}]<\sqrt{1-|\gamma_{\mathbf{k}}|^{2}}-1$, 
the two-€"mode generalized coherent state $\ket{r_{\mathbf{k}}, 
\phi_{\mathbf{k}}}$ is a two-€"mode squeezed state with mean variance being 
the associated EPR-uncertainty \cite{giedke03}. For 
$\Delta(r_{\mathbf{k}} ,  \phi_{\mathbf{k}})\geqslant1$, on the other hand, the 
EPR-uncertainty is constant and equal to $1$. In the latter case, the amount of 
nonlocal correlations vanishes \cite{giedke03} although the magnon CV entanglement 
is nontrivial; thus, the magnon CV entanglement can only be related to the EPR-uncertainty 
in the squeezing domain.

The relation in Eq. \eqref{AMV} allows one 
to evaluate the CV entanglement in terms of $\Delta(r_{\mathbf{k}}, \phi_{\mathbf{k}})$.
For real $\gamma_{\mathbf{k}}$, which correspond to $\phi_{\mathbf{k}}=0$ or $\pi$, we find   
\begin{eqnarray}
E^{(a, b)}_{0}& = & 
\left[\frac{1+\Delta(r_{\mathbf{k}} ,  \phi_{\mathbf{k}})}{2\sqrt{\Delta(r_{\mathbf{k}},  
\phi_{\mathbf{k}})}}\right]^{2} 
\log_2 \left[ \frac{1+\Delta(r_{\mathbf{k}},  \phi_{\mathbf{k}})}{2\sqrt{\Delta(r_{\mathbf{k}},  
\phi_{\mathbf{k}})}} 
\right]^2
\nonumber \\
 & & - \left[ \frac{1-\Delta(r_{\mathbf{k}},  \phi_{\mathbf{k}})}{2\sqrt{\Delta(r_{\mathbf{k}},  
 \phi_{\mathbf{k}})}}
\right]^2 \log_2 \left[ 
\frac{1-\Delta(r_{\mathbf{k}},  \phi_{\mathbf{k}})}{2\sqrt{\Delta(r_{\mathbf{k}},  \phi_{\mathbf{k}})}}
\right]^2 .\ \ \ \ \ \ 
\end{eqnarray}
The parameter $\Delta(r_{\mathbf{k}}, \phi_{\mathbf{k}})$, which depends on the lattice 
geometry and the choice of $\mathbf{k}$, can be accessed experimentally by detecting 
coherences of quantum fields, the quadratures, with homodyne detection techniques 
\cite{gross2011, peise2015} adapted to a possible magnon-photon coupling 
\cite{yuan2017,lachance20}. This may be an avenue forward for experimental detection 
of the magnon CV entanglement.

In order to further explore the material-specific features of magnon entanglement, we 
consider a more general spin Hamiltonian, that also has Dzyaloshinskii-Moriya interaction,
\begin{eqnarray}
H=H_{0}+H_{\text{DM}}
\label{H0DM}
\end{eqnarray}
with $H_{\text{DM}} =  \sum_{\langle ij \rangle} \mathbf{D}_{ij}\cdot\left( 
\mathbf{S}_{i}\times\mathbf{S}_{j}\right)$ being the DM term 
with $\mathbf{D}_{ij}=-\mathbf{D}_{ji}$ pointing along the same fixed direction $\mathbf{D}$ 
for all nearest 
neighbor spin pairs. By assuming $D=|\mathbf{D}|$, $H$ takes the form
\begin{eqnarray}
H&  =& \sum_{\mathbf{k}}\epsilon_{\mathbf{k}} 
(\alpha_{\mathbf{k}}^{\dagger}\alpha_{\mathbf{k}} + 
\beta_{\mathbf{k}}^{\dagger}\beta_{\mathbf{k}}) 
\nonumber\\
  & &+ izDS\sum_{\mathbf{k}}(\gamma_{\mathbf{k}} \alpha_{\mathbf{k}}
\beta_{\mathbf{k}} -\gamma_{-\mathbf{k}} \alpha_{\mathbf{k}}^{\dagger} 
\beta_{\mathbf{k}}^{\dagger}),
\end{eqnarray}
which is not diagonal anymore in the $(\alpha,\beta)$  modes. 
Without loss of generality, we assume real-valued $u_{\mathbf{k}}$ 
in the Bogoliubov transformation of Eq. \eqref{eq:FBT}.
The second summation of off-diagonal terms 
on the right hand side implies that there is mixing between $\alpha$ 
and $\beta$ modes in the presence of the DM
interaction. This may cause extra magnon CV entanglement in the ground state of the 
system. To see this, we diagonalize $H$ by applying another Bogoliubov transformation 
 \begin{eqnarray}
\left(
\begin{array}{cc}
  \alpha_{\mathbf{k}}    \\
   \beta_{\mathbf{k}}^{\dagger}       
\end{array}
\right)=\left(
\begin{array}{cc}
  \eta_{\mathbf{k}}& \zeta_{\mathbf{k}}    \\
\zeta^{*}_{\mathbf{k}}& \eta^{*}_{\mathbf{k}}       
\end{array}
\right)\left(
\begin{array}{cc}
  \tilde{\alpha}_{\mathbf{k}}    \\
   \tilde{\beta}_{\mathbf{k}}^{\dagger}       
\end{array}
\right),
\label{BT2}
\end{eqnarray}
where $\zeta_{\mathbf{k}}$ and $\eta_{\mathbf{k}}$ are given by
 \begin{eqnarray}
|\eta_{\mathbf{k}}|^{2} =  \frac{1}{2\sqrt{1-|\Gamma_{\mathbf{k}}|^{2}}}+\frac{1}{2},&&\ \ \ \ 
|\zeta_{\mathbf{k}}|^{2} = \frac{1}{2\sqrt{1-|\Gamma_{\mathbf{k}}|^{2}}} - \frac{1}{2} , 
\nonumber\\
\frac{\zeta_{\mathbf{k}}}{\eta_{\mathbf{k}}^{*}} = - & & 
\frac{1-\sqrt{1-
|\Gamma_{\mathbf{k}}|^{2}}}{\Gamma_{\mathbf{k}}},
\label{BTC}
\end{eqnarray}
provided $\Gamma_{\mathbf{k}} = 
\frac{iD\gamma_{\mathbf{k}}}{J\sqrt{1-|\gamma_{\mathbf{k}}|^{2}}}$ with 
$|\Gamma_{\mathbf{k}}|<1$. In the $(\tilde{\alpha}_{\mathbf{k}},\tilde{\beta}_{\mathbf{k}}$) 
modes, the Hamiltonian $H$ takes the diagonal form
\begin{eqnarray}
H & = & 
\sum_{\mathbf{k}}\tilde{\epsilon}_{\mathbf{k}}
(\tilde{\alpha}_{\mathbf{k}}^{\dagger}
\tilde{\alpha}_{\mathbf{k}} + \tilde{\beta}_{\mathbf{k}}^{\dagger}
\tilde{\beta}_{\mathbf{k}}) 
\label{GDH}
\end{eqnarray}
with the dispersion relation ${\epsilon}_{\mathbf{k}} = 
zS \sqrt{J^{2}(1-|\gamma_\mathbf{k}|^{2}) - 
D^{2}|\gamma_\mathbf{k}|^{2}}$ (see Supplemental Material for derivation). 

\begin{figure}[h]
\begin{center}
\includegraphics[width=75mm]{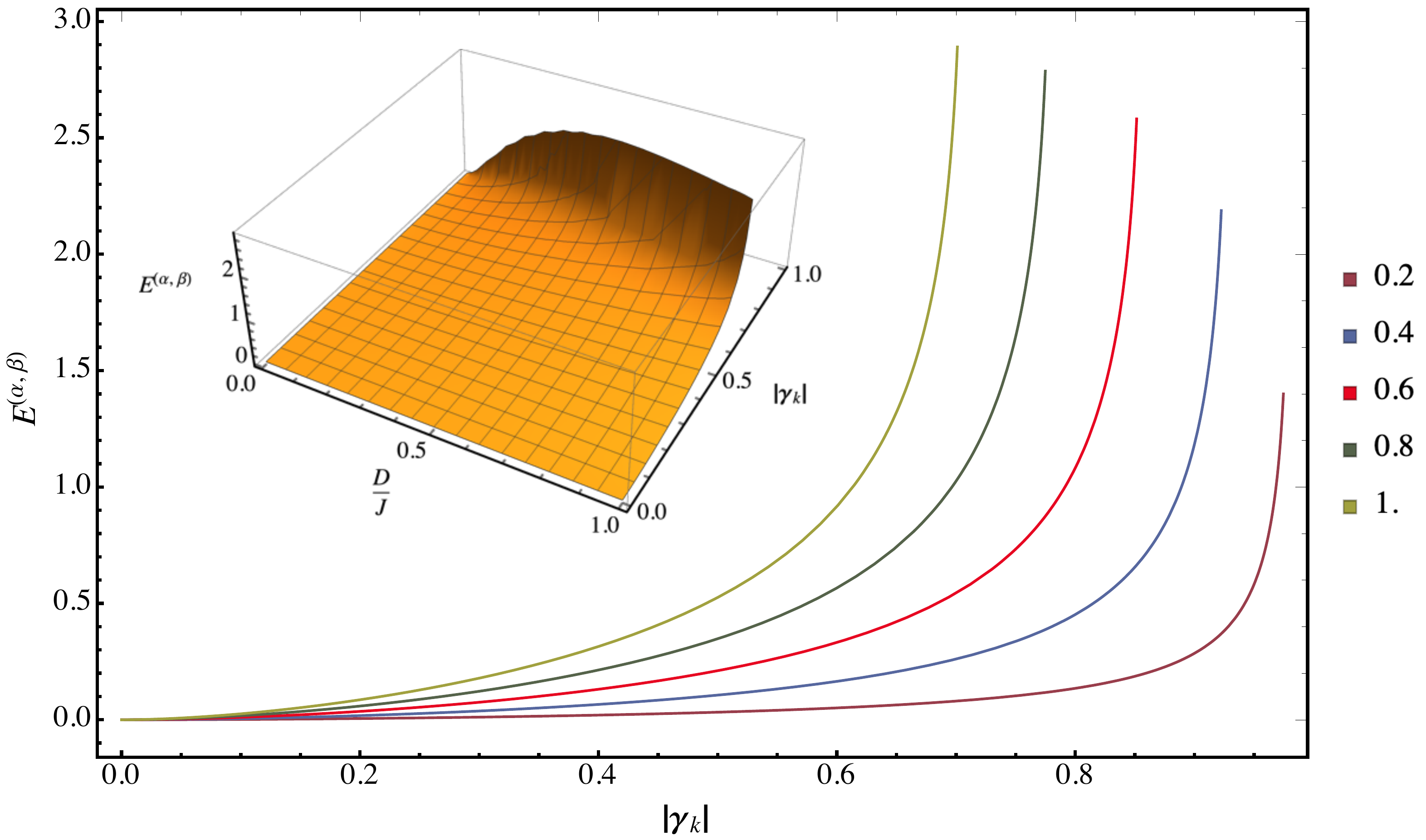}
\end{center}
\caption{(Color online). The entropy of entanglement $E^{(\alpha, \beta)}$ of the two-€"mode 
generalized coherent state $\ket{\tilde{r}_{\mathbf{k}}, \tilde{\phi}_{\mathbf{k}}}$ as a function 
of $\left| \gamma_\mathbf{k} \right|$ and $\frac{D}{J}$ in the $(\alpha,\beta)$ modes. In the 
main figure, we show plots for different values of $\frac{D}{J}$, while the inset is a 
three-dimensional plot of the entropy of entanglement as a function of $\frac{D}{J}$ 
and $\left| \gamma_\mathbf{k} \right|$.}
\label{fig:EEDMP}
\end{figure}

The ground state of the diagonal Hamiltonian is a product state $\ket{\psi} = \prod_{\mathbf{k}} 
\ket{0;\tilde{\alpha}_{\mathbf{k}}} \ket{0; \tilde{\beta}_{\mathbf{k}}}$, where 
$ \ket{0; \tilde{\alpha}_{\mathbf{k}}}$ and $\ket{0; 
\tilde{\beta}_{\mathbf{k}}}$ are vacuum states of 
$\tilde{\alpha}_{\mathbf{k}}$ and $\tilde{\beta}_{\mathbf{k}}$, respectively. In this basis, the 
magnon entanglement is absent.
Using the inverse transformation back into the $(\alpha,\beta)$ modes, 
we express the ground state as
\begin{eqnarray}
\ket{\psi}=\prod_{\mathbf{k}} \ket{\tilde{r}_{\mathbf{k}}, 
\tilde{\phi}_{\mathbf{k}}} 
\end{eqnarray}
with the entangled two-€"mode generalized coherent state
\begin{eqnarray}
\ket{\tilde{r}_{\mathbf{k}}, \tilde{\phi}_{\mathbf{k}}} = 
\frac{1}{\cosh\tilde{r}_{\mathbf{k}}} \sum_{n=0}^{\infty} 
\tanh^{n}\tilde{r}_{\mathbf{k}}e^{in\tilde{\phi}_{\mathbf{k}}}
\ket{n; \alpha_{\mathbf{k}}}\ket{n; \beta_{\mathbf{k}}} ,
\nonumber\\
\label{TMSS}
\end{eqnarray}
where $\ket{n; \alpha_{\mathbf{k}}}$ and $\ket{n; \beta_{\mathbf{k}}}$ are the 
$n$th excitation of $\alpha_{\mathbf{k}}$ and $\beta_{\mathbf{k}}$,  respectively. 
Here, $\tilde{r}_{\mathbf{k}}$ and  $\tilde{\phi}_{\mathbf{k}}$ are specified by 
$e^{i\tilde{\phi}_{\mathbf{k}}}\tanh\tilde{r}_{\mathbf{k}} =
\frac{\zeta_{\mathbf{k}}}{\eta_{\mathbf{k}}^{*}}$ with 
$\tilde{r}_{\mathbf{k}}\equiv \tilde{r}_{\mathbf{k}}(\gamma_\mathbf{k}, \frac{D}{J})\geqslant 0$, 
and $\tilde{\phi}_{\mathbf{k}}\equiv \tilde{\phi}_{\mathbf{k}}(\gamma_\mathbf{k}, \frac{D}{J}) = 
\pi-\arg[\Gamma_{\mathbf{k}}]=\frac{\pi}{2}-\arg[\gamma_\mathbf{k}]$. In the case of $D=0$, 
the only relevant term is $n=0$, i.e., $\ket{\tilde{r}_{\mathbf{k}}, \tilde{\phi}_{\mathbf{k}}} =
\ket{0; \alpha_{\mathbf{k}}}\ket{0; \beta_{\mathbf{k}}}$ and thus $\ket{\psi}=\ket{\psi_0}$.
The entropy of entanglement in the $(\alpha, \beta)$ modes
\begin{eqnarray}
E^{(\alpha, \beta)}=|\eta_\mathbf{k}|^{2}\log_2 |\eta_\mathbf{k}|^{2}-|\xi_\mathbf{k}|^{2} 
\log_2 |\xi_\mathbf{k}|^{2},
\label{EEDM}
\end{eqnarray}
is a function of $\left| \gamma_\mathbf{k} \right|$ and the relative 
coupling strength $\frac{D}{J}$. Fig. \ref{fig:EEDMP} shows 
magnon CV entanglement $E^{(\alpha, \beta)}$ as a function of $\left| \gamma_\mathbf{k} \right|$ 
for selected values of  $\frac{D}{J}$. 

In the $(\alpha, \beta)$ 
modes, the two-mode magnon entanglement, Eq.~\eqref{EEDM}, is non-trivial, while, 
as shown above, the symmetric Heisenberg interaction $H_0$ on its own does not 
generate any magnon entanglement in these modes, i.e., $E^{(\alpha, \beta)}_{0}=0$. Thus, 
the antisymmetric DM interaction is mainly responsible for the entanglement contribution in 
Eq.~\eqref{EEDM}, and we identify $E^{(\alpha, \beta)}_{\text{DM}}=E^{(\alpha, \beta)}$ as 
the DM-induced entanglement. 

\begin{figure}[h]
\begin{center}
\includegraphics[width=75mm]{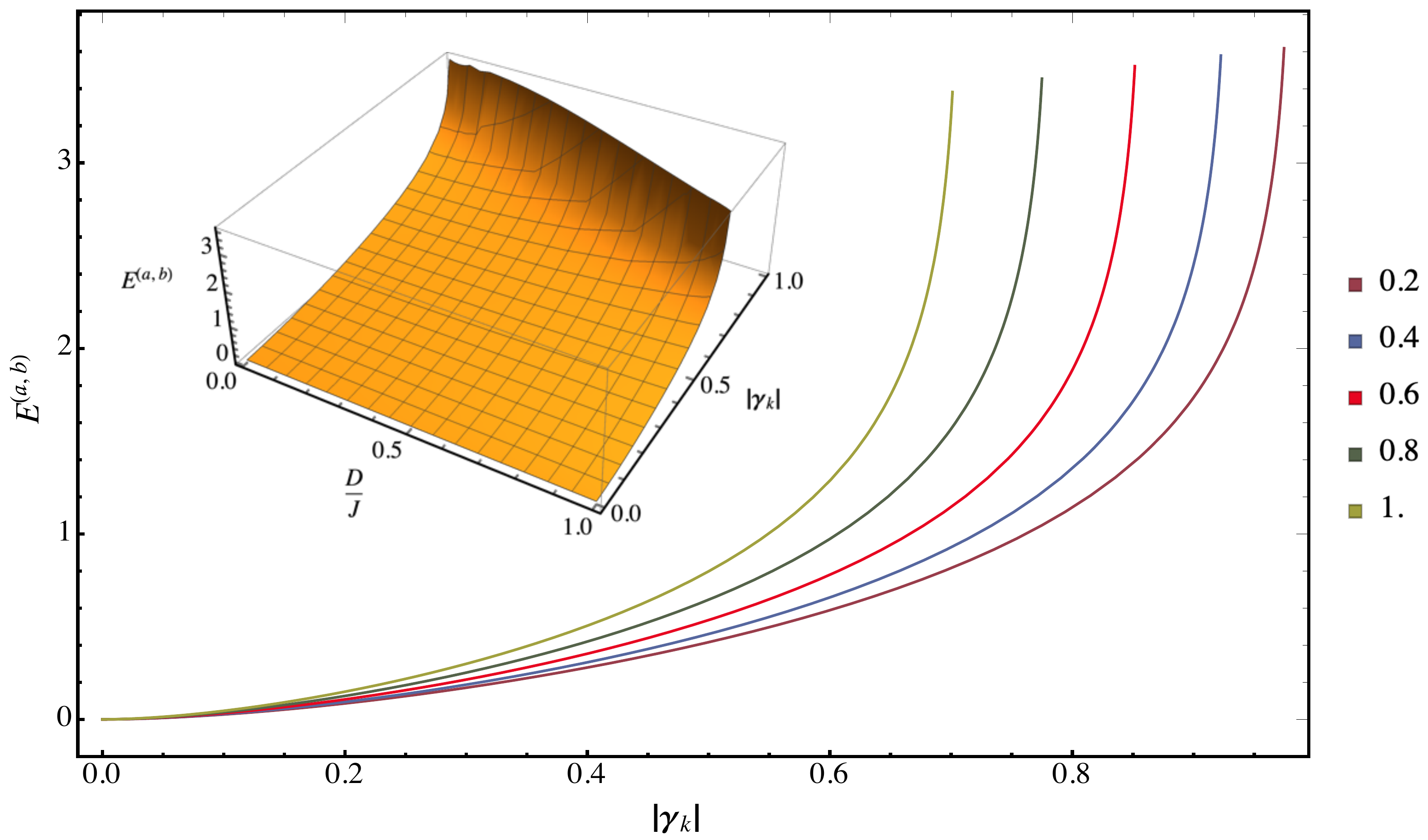}
\end{center}
\caption{(Color online). The entropy of entanglement of the two-€"mode generalized 
coherent state $\ket{\hat{r}_{\mathbf{k}}, \hat{\phi}_{\mathbf{k}}}$ as a function of 
$\left|\gamma_\mathbf{k}\right|$ and the relative coupling strength $\frac{D}{J}$ in 
the $(a, b)$ modes. In the main figure, we show several sections of this function for 
different values of $\frac{D}{J}$, and the inset is a full three-dimensional plot.}
\label{fig:GSTE}
\end{figure}

To have a clearer picture of the hierarchy of the magnon CV entanglement, 
we transform the ground state $\ket{\psi}$ back into the original 
$(a, b)$ modes
 \begin{eqnarray}
\ket{\psi}=\prod_{\mathbf{k}} \ket{\hat{r}_{\mathbf{k}}, \hat{\phi}_{\mathbf{k}}},
\end{eqnarray}
with the  two-€"mode generalized coherent state 
\begin{eqnarray}
\ket{\hat{r}_{\mathbf{k}}, \hat{\phi}_{\mathbf{k}}}=\frac{1}{\cosh\hat{r}_{\mathbf{k}}} 
\sum_{n=0}^{\infty}\tanh^{n}\hat{r}_{\mathbf{k}}e^{in\hat{\phi}_{\mathbf{k}}} 
\ket{n; a_{\mathbf{k}}}\ket{n; b_{\mathbf{k}}},\ \ \ \ \ \ 
\label{TS}
\end{eqnarray}
where
$e^{i\hat{\phi}_{\mathbf{k}}}\tanh\hat{r}_{\mathbf{k}}=\frac{v_{\mathbf{k}} 
\eta^{*}_{\mathbf{k}}+u_{\mathbf{k}}\zeta_{\mathbf{k}}}{u^{*}_{\mathbf{k}} 
\eta^{*}_{\mathbf{k}}+v^{*}_{\mathbf{k}}\zeta_{\mathbf{k}}}$, $\hat{r}_{\mathbf{k}} \equiv 
\hat{r}_{\mathbf{k}}(\gamma_\mathbf{k}, \frac{D}{J})\geqslant 0$, and 
$\hat{\phi}_{\mathbf{k}}\equiv \hat{\phi}_{\mathbf{k}}(\gamma_\mathbf{k}, \frac{D}{J}) = 
\pi-\arg[\gamma_\mathbf{k}(1+i\frac{D}{J})]$. The total entropy of entanglement in the 
$(a, b)$ modes for this state is  
\begin{eqnarray}
E^{(a, b)}&=&\cosh^{2}\hat{r}_{\mathbf{k}}\log_2[\cosh^{2}\hat{r}_{\mathbf{k}}] - 
\sinh^{2}\hat{r}_{\mathbf{k}}\log_2[\sinh^{2}\hat{r}_{\mathbf{k}}].\ \ \ \ \ \ \ 
\label{TE}
\end{eqnarray}
Figure~\ref{fig:GSTE} shows the entanglement of Eq.~\eqref{TE} as function of 
$\left| \gamma_\mathbf{k} \right|$, for selected values of $\frac{D}{J}$. Note that 
increasing values of the DM interaction leads to an enhancement of the entanglement 
for all $\left| \gamma_\mathbf{k} \right|$.

Since $E^{(a, b)}|_{D=0}=E_0^{(a, b)}$, we may write
\begin{eqnarray}
E^{(a, b)}=E_{0}^{(a, b)}+E^{(a, b)}_{\text{DM}},
\end{eqnarray}
where $E^{(a, b)}_{\text{DM}}$ 
vanishes in the absence of DM interaction, and we refer to it as the DM-induced 
entanglement in the $(a,b)$ modes. Unlike the $(\alpha, \beta)$ modes, in the $(a, b)$ 
modes both the Heisenberg and the DM interactions induce non-zero contributions to 
the magnon entanglement in the ground state. 

The magnon CV entanglement is an intrinsic property of antiferromagnets that depends 
on the geometry of the spin lattice as encoded in $\gamma_\mathbf{k}$ and on the relative 
coupling strength $\frac{D}{J}$. Both parameters are material-dependent and can vary 
strongly from system to system. This opens an interesting route to search for suitable 
entanglement hosts among the existing thousands of magnetic compounds, and poses 
a natural question of how magnon entanglement can be detected in an experiment. 
Similar to what was discussed above in the pure Heisenberg case, the magnon CV 
entanglement in the presence of DM interaction can be measured experimentally by 
detecting quadratures corresponding to the arithmetic mean variance 
$\Delta(\hat{r}_{\mathbf{k}}, \hat{\phi}_{\mathbf{k}})$ in a homodyne detection setup 
\cite{gross2011, peise2015} adapted to a possible magnon-photon coupling 
\cite{yuan2017,lachance20}. In the $(a, b)$ modes, one may extract the parameter 
$\hat{r}_{\mathbf{k}}$ from 
\begin{equation}
\Delta(\hat{r}_{\mathbf{k}},  \hat{\phi}_{\mathbf{k}})=\cosh 2\hat{r}_{\mathbf{k}} - 
\sinh 2\hat{r}_{\mathbf{k}}\cos\hat{\phi}_{\mathbf{k}}
\end{equation}
to evaluate the total entanglement $E^{(a, b)}$ in Eq.~(\ref{TE}). Here, we used 
Eq.~(\ref{AMV}) replacing the state 
$\ket{r_{\mathbf{k}}, \phi_{\mathbf{k}}}$ by the state $\ket{\hat{r}_{\mathbf{k}}, 
\hat{\phi}_{\mathbf{k}}}$. In the domain of $\tanh\hat{r}_{\mathbf{k}} \leqslant 
\cos\hat{\phi}_{\mathbf{k}}$ corresponding to $\mathrm{Re}[\gamma_{\mathbf{k}} 
(J+iD)]<\sqrt{J^{2}(1-|\gamma_{\mathbf{k}}|^{2})-|\gamma_{\mathbf{k}}|^{2}D^{2}}-J$, 
the two-€"mode generalized coherent state $\ket{\hat{r}_{\mathbf{k}}, \hat{\phi}_{\mathbf{k}}}$ 
is also a two-€"mode squeezed state, and the mean variance 
$\Delta(\hat{r}_{\mathbf{k}},  \hat{\phi}_{\mathbf{k}})$ is the associated
EPR-uncertainty \cite{giedke03}.

\begin{figure}[h]
\begin{center}
\includegraphics[width=80mm]{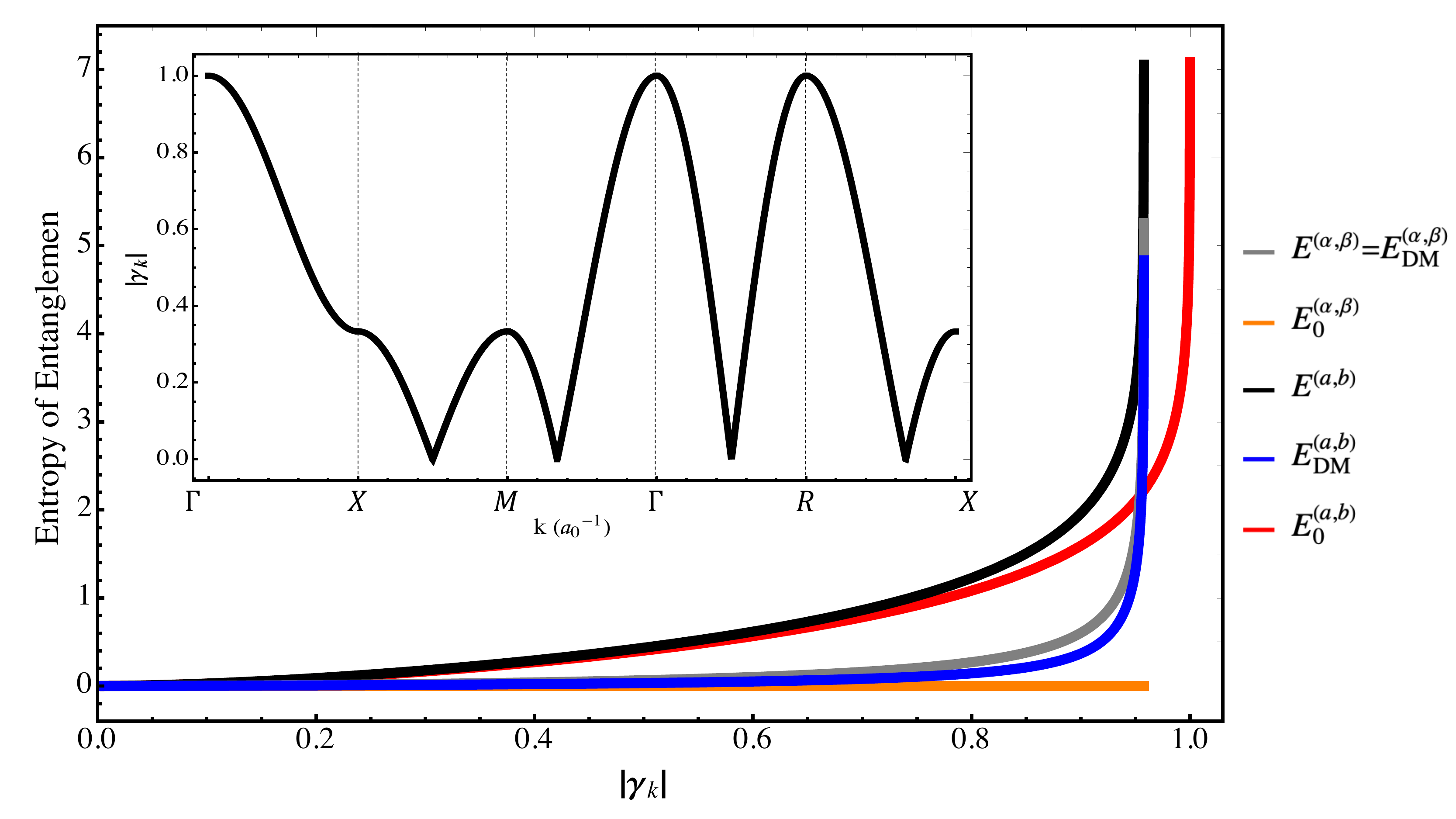}
\end{center}
\caption{(Color online). Hierarchy of magnon entanglement for an antiferromagnet as 
a function of $\left| \gamma_{\mathbf{k}} \right|$. Definition of the different entanglement 
entropies is given in the main text. An effective DM interaction with the strength of 
$30\%$ of the Heisenberg exchange $J$ is used. The inset illustrates the dependence 
of entanglement on $\mathbf{k}$ in a simple cubic crystal structure.}
\label{fig:HoE}
\end{figure}

We conclude with some final remarks.
In the analysis of different bosonic modes, we notice different types of two-mode magnon 
entanglement residing in the ground state. 
In Fig.~\ref{fig:HoE}, we compare entropies of entanglement for an antiferromagnet with 
a simple cubic crystal structure, where the Hamiltonian
is effectively described by nearest neighbor Heisenberg exchange as well as DM interaction 
with typical ratio $D/J\approx 30\%$ \cite{meyer2017, chern2006, zakeri2017}. It can be seen 
that from $(a, b)$ modes to $(\alpha, \beta)$ modes the Heisenberg contribution to 
entanglement decreases while the DM-induced magnon CV entanglement increases. 
This is due to the fact that different bosonic modes represent different tensor product 
structures of the Hilbert space \cite{zanardi2004}. While the $(a,b)$ modes describe 
naturally identifiable magnon modes, being associated with each sublattice, the 
$(\alpha, \beta)$ and  $(\tilde{\alpha},\tilde{\beta})$ modes are hybridized and their 
number states are described by superpositions of excitations in the $(a, b)$ modes. 
Although the stronger entanglement, a feature particularly useful in quantum information 
science and technology, is available in the $(a, b)$ modes, the usefulness of these modes 
in an actual experiment must be determined by a suitable tensor product decomposition.

The condition for diagonalizing the Hamiltonians in terms of bosonic operators is that 
$|\gamma_{\mathbf{k}}|<1$ in the pure Heisenberg case and $|\gamma_{\mathbf{k}}|^{2} < 
\frac{J^{2}}{J^{2}+D^{2}}$ in the presence of DM interaction.
Since ${D}$ is typically less than a few tenths of ${J}$ for most materials 
\cite{meyer2017, chern2006, zakeri2017}, this condition is not satisfied only for a 
very small part of the BZ, e.g., the region around zone center. 
We would like to remark that the entire BZ may be included in this analysis by considering a 
Hamiltonian that possesses single ion uniaxial anisotropy, $-\mathcal{K}({\bf n} \cdot {\bf S})^{2}$, 
e.g., with easy-axis ${\bf n}$ along the direction of the DM vector, as long as  
$|1+\frac{2\mathcal{K}}{zJ}| > \sqrt{1+\frac{D^{2}}{J^{2}}}$. 
Note that any change of symmetry in the Hamiltonian 
introduces new magnon modes and hence new levels of 
entanglement contribution in the hierarchy of two-mode 
magnon entanglement. Our general message is not changed 
by this, although the technical level of calculations 
may become more intricate. It is interesting to note 
that already very mild uniaxial anisotropy of $0.001\%$ 
of $J$ along $z$ axis, when included in a pure 
Heisenberg Hamiltonian ($D=0$), allows to 
regularize the magnon CV entanglement dependence 
on $|\gamma_{\bf k}|$. At $|\gamma_{\bf k}|=1$, we 
obtain $E_{0}^{(a, b)}\approx 9.04$. In a more generic case of Heisenberg-DMI with 
$D/J=0.1$, $z=6$ and uniaxial anisotropy of $\mathcal{K}/J=$0.015 along $\mathbf{D}$, 
we find $E^{(a, b)}\approx 8.094$ and $E^{(\alpha, \beta)}\approx 4.766$ at 
$|\gamma_{\bf k}|=1$.

In contrast to Ref.~\cite{bossini2019}, where photoinduced spin dynamics was employed 
to trigger entanglement between a pair of magnon modes, our analysis shows that ground 
state two-mode magnon entanglement in antiferromagnets is an intrinsic property of the 
magnetic structure that 
is already given by the geometry of the spin lattice and exchange 
couplings, which should be 
accessible to experimental detection.
We have examined the magnon entanglement in quantum magnetic structures 
with nearest neighbor antiferromagnetic Heisenberg exchange and DM interaction.
The analysis is appropriate for many classes of compounds, but we would like to mention 
in particular the transition metal oxides that have a vast crystallographic phase space, 
which allows both for tunability of $\frac{D}{J}$ as well as $\gamma_\mathbf{k}$.
A concrete example that is known to exhibit only nearest neighbor Heisenberg exchange 
is SrMnO$_3$ \cite{zhu2020}. We also note that materials like  La$_2$CuO$_4$ 
\cite{voigt1996}, FeBO$_3$, and CoCO$_3$ \cite{beutier2017} are well studied 
antiferromagnets that are known to have DM interaction in the here studied range 
of $\frac{D}{J}$.

\begin{acknowledgments}
{\it Acknowledgments} --The authors acknowledge financial support from Knut and 
Alice Wallenberg Foundation 
through Grant No. 2018.0060. O.E.  acknowledges support from eSSENCE, SNIC and 
the Swedish Research Council (VR). D.T. acknowledges support from the Swedish Research 
Council (VR) through Grant No. 2019-03666. A.B. acknowledges financial support from the 
Russian Science Foundation through Grant No. 18-12-00185.
A.D. acknowledges financial support from the Swedish Research Council (VR) through 
Grants No. 2015-04608,  2016-05980, and VR 2019-05304. E.S. acknowledges financial 
support from the Swedish Research Council (VR) through Grant No. 2017-03832. 
Some of the computations were performed on resources provided by the Swedish 
National Infrastructure for Computing (SNIC) at the National Supercomputer Center (NSC), 
Link\"oping University, the PDC Centre for High Performance Computing (PDC-HPC), KTH, 
and the High Performance Computing Center North (HPC2N), Ume{\aa} University.
\end{acknowledgments}


\begin{thebibliography}{99}
\bibitem{remark1} Entropy of entanglement $E_{\psi^{AB}}$ of a pure bipartite state 
$\ket{\psi^{AB}}$ is defined as the von Neumann entropy of the reduced density operator 
of one of the two subsystems $A$ or $B$. Explicitly, given the Schmidt form 
$\ket{\psi^{AB}} = \sum_n c_n \ket{\varphi_n^A}\ket{\varphi_n^B}$, one 
has $E_{\psi^{AB}} = - \sum_n |c_n|^2 \log_2 |c_n|^2$.
\bibitem{note1} Dimensionless position and momentum quadratures of the
$a_{\mathbf{k}}$ and $b_{\mathbf{k}}$ modes are $X_{\mathbf{k}}^{A} = 
\frac{a_{\mathbf{k}} + a_{\mathbf{k}}^{\dagger}}{\sqrt{2}}$, 
$X_{\mathbf{k}}^{B} = 
\frac{b_{\mathbf{k}}+b_{\mathbf{k}}^{\dagger}}{\sqrt{2}}$, 
$P_{\mathbf{k}}^{A} = 
\frac{a_{\mathbf{k}}-a_{\mathbf{k}}^{\dagger}}{i\sqrt{2}}$, 
and $P_{\mathbf{k}}^{B} = \frac{b_{\mathbf{k}} - 
b_{\mathbf{k}}^{\dagger}}{i\sqrt{2}}$.
\end{thebibliography}
\end{document}